\documentclass[conference]{IEEEtran}
\usepackage[bookmarksnumbered,bookmarksopen,unicode,colorlinks=true,urlcolor=blue,citecolor=blue,linkcolor=blue]{hyperref}
\usepackage{orcidlink}

\newcommand{\upp}{\vspace*{-0.5em}}

\usepackage{cite}
\ifCLASSINFOpdf
\else
\fi
\usepackage{amsmath,amssymb}
\usepackage{physics}
\usepackage{mathtools}
\usepackage{algorithm}
\usepackage{algpseudocode}

\usepackage[caption=false,font=normalsize,labelfont=sf,textfont=sf]{subfig}

\usepackage{xcolor}

\newif\ifdraft
\drafttrue
\ifdraft

\newcommand{\linebreakand}{%
  \end{@IEEEauthorhalign}
  \hfill\mbox{}\par
  \mbox{}\hfill\begin{@IEEEauthorhalign}
}

\newcommand{\maxcut}{{\scshape 
MaxCut
}}

\begin{document}
\title{Quantum and Classical Correlations in\\Shrinking Algorithms for Optimization}



% \author{\IEEEauthorblockN{Michael Shell\IEEEauthorrefmark{1},
% Homer Simpson\IEEEauthorrefmark{2},
% James Kirk\IEEEauthorrefmark{3}, 
% Montgomery Scott\IEEEauthorrefmark{3} and
% Eldon Tyrell\IEEEauthorrefmark{4}}
% \IEEEauthorblockA{\IEEEauthorrefmark{1}School of Electrical and Computer Engineering\\
% Georgia Institute of Technology,
% Atlanta, Georgia 30332--0250\\ Email: see http://www.michaelshell.org/contact.html}
% \IEEEauthorblockA{\IEEEauthorrefmark{2}Twentieth Century Fox, Springfield, USA\\
% Email: homer@thesimpsons.com}
% \IEEEauthorblockA{\IEEEauthorrefmark{3}Starfleet Academy, San Francisco, California 96678-2391\\
% Telephone: (800) 555--1212, Fax: (888) 555--1212}
% \IEEEauthorblockA{\IEEEauthorrefmark{4}Tyrell Inc., 123 Replicant Street, Los Angeles, California 90210--4321}}

\author{\IEEEauthorblockN{
Victor Fischer\IEEEauthorrefmark{6}\IEEEauthorrefmark{2}\IEEEauthorrefmark{7}\IEEEauthorrefmark{3}\orcidlink{0009-0003-2976-2201}, 
Maximilian Passek\IEEEauthorrefmark{1}\IEEEauthorrefmark{2}\IEEEauthorrefmark{7}\IEEEauthorrefmark{3},
Friedrich Wagner\IEEEauthorrefmark{4}\IEEEauthorrefmark{5}\orcidlink{0000-0001-9031-1870},
Jernej Rudi Fin\v zgar\IEEEauthorrefmark{1}\IEEEauthorrefmark{2}\orcidlink{0000-0003-4393-827X},
Lilly Palackal\IEEEauthorrefmark{6}\IEEEauthorrefmark{2}\orcidlink{0009-0007-4858-7356},\\
Christian B.~Mendl\IEEEauthorrefmark{2}\orcidlink{0000-0002-6386-0230}}

\IEEEauthorblockA{
                  \IEEEauthorrefmark{6}Infineon Technologies AG, Munich, Germany\\
                  \IEEEauthorrefmark{1}BMW Group, Munich, Germany\\
                  \IEEEauthorrefmark{4}University of Erlangen-Nuremberg, Germany\\
                  \IEEEauthorrefmark{5}Fraunhofer Institute for Integrated Circuits, Erlangen, Germany\\
                  \IEEEauthorrefmark{2}Technical University of Munich, Germany\\
                  \IEEEauthorrefmark{7}Ludwig Maximilian University of Munich, Germany\\
                  \IEEEauthorrefmark{3}{\footnotesize Authors contributed equally}\upp\upp\upp}}


\maketitle
\begin{abstract}
Understanding the benefits of quantum computing for solving combinatorial optimization problems (COPs) remains an open research question.
In this work, we extend and analyze algorithms that solve COPs by recursively shrinking them.
The algorithms leverage correlations between variables extracted from quantum or classical subroutines to recursively simplify the problem.
We compare the performance of the algorithms equipped with correlations from the quantum approximate optimization algorithm (QAOA) as well as the classical linear programming (LP) and semi-definite programming (SDP) relaxations.
This allows us to benchmark the utility of QAOA correlations against established classical relaxation algorithms.
We apply the recursive algorithm to \maxcut\ problem instances with up to a hundred vertices at different graph densities.
Our results indicate that LP outperforms all other approaches for low-density instances, while SDP excels for high-density problems.
Moreover, the shrinking algorithm proves to be a viable alternative to established methods of rounding LP and SDP relaxations.
In addition, the recursive shrinking algorithm outperforms its bare counterparts for all three types of correlations, i.e., LP with spanning tree rounding, the Goemans-Williamson algorithm, and conventional QAOA.
While the lowest depth QAOA consistently yields worse results than the SDP, our tensor network experiments show that the performance increases significantly for deeper QAOA circuits.
\end{abstract}

\IEEEpeerreviewmaketitle

\section{Introduction}
Combinatorial optimization problems (COPs)
are highly relevant in both science and industry.
However, they are generally challenging to solve efficiently because of the exponential growth of the solution space with increasing problem size.
Traditionally, a variety of classical strategies, including linear programming (LP)~\cite{Raghavan1887} and semi-definite programming (SDP) relaxations~\cite{goemans1995improved}, are applied to find approximate solutions to COPs.
Recently, approaches utilizing quantum resources have emerged as promising alternative solution strategies~\cite{Abbas}.
Among these, the quantum approximate optimization algorithm (QAOA)~\cite{farhi2014quantum} is particularly interesting because of its universality and adaptive complexity.
QAOA is a hybrid quantum-classical variational algorithm designed to find approximate solutions to unconstrained binary optimization problems using noisy-intermediate scale quantum (NISQ) devices and beyond.

It is unclear whether NISQ algorithms can provide improvements over classical methods for solving relevant COPs in industry.
In fact, several issues of the near-term variational approaches have been identified, such as barren plateaus~\cite{bittel2021} and noise of quantum devices~\cite{stilck_franca_limitations_2021}.
Another restriction of QAOA is due to its \emph{locality}, i.e., the effect that only qubits separated by less than a certain distance in the graph representation of a problem can interact with each other at a given circuit depth.
This property implies limitations on the performance of QAOA~\cite{bravyi_obstacles_2020, farhi_quantum_2020, farhi_quantum_2020-1,chou2022limitations}.

The recursive QAOA (RQAOA) was introduced by Bravyi~\emph{et al.}~\cite{bravyi_obstacles_2020, bravyi_hybrid_2022} to overcome the locality-induced limitations of QAOA.
The algorithm operates by recursively simplifying the problem based on correlations between variables obtained from QAOA.
The two steps of computing the correlations and fixing the variables are executed iteratively until the problem is fully solved.
By introducing new connections between previously unlinked variables, RQAOA overcomes the locality of QAOA.
Several works have generalized and extended RQAOA.
These proposals include using problem-specific update rules~\cite{finzgar_quantum-informed_2024, brady_iterative_2023}, analog quantum devices~\cite{finzgar_quantum-informed_2024} or a shrinking procedure based on the classical calculation of correlations~\cite{wagner_enhancing_2023}.
It has been shown that those recursive algorithms outperform the original QAOA for many problem instances~\cite{Bae_2024, bravyi_hybrid_2022, bravyi2021classical}.
However, it is not clear whether the improved results are due to the recursive shrinking procedure or the quantum correlations.
Here, we aim to give insights into this question.

\paragraph*{Contributions}
Our study evaluates how different classical and quantum methods for calculating correlations affect the performance of a shrinking algorithm similar to RQAOA.
To this end, we introduce new approaches of calculating correlations, beyond the known QAOA, through classical means. 
The investigated routines for calculating correlations are the classical LP and SDP relaxations, and QAOA.
Our numerical experiments on instances of \maxcut allow us to benchmark the utility of quantum correlations against correlations obtained from established classical approximation algorithms.
Moreover, the SDP shrinking procedure introduced in this work is a novel recursive approximation algorithm inspired by the Goemans-Williamson approach, specifically designed for \maxcut problems. It outperforms the original Goemans-Williamson algorithm for the randomized problem instances considered in this study. 
Furthermore, we show that the shrinking algorithm employed here is a viable alternative to traditional rounding routines for LP and SDP relaxations.

The remainder of the paper is organized as follows.
In Section~\ref{sec: methods} we introduce the \maxcut problem and the different routines (LP, SDP and QAOA) for calculating correlations.
In addition, we present the shrinking algorithm that utilizes the calculated correlations.
In Section~\ref{sec: results} we analyze the performance of the shrinking algorithm equipped with different correlation sources before discussing the implications of our findings in Section~\ref{sec: discussion}.
In Section~\ref{sec: outlook} we conclude by suggesting potential future research directions.

\section{Methods}\label{sec: methods}
In this section, we first introduce the \maxcut problem.
Next, we explain how to compute the correlations using the LP, SDP, and QAOA.
Finally, we present the shrinking algorithm that uses the correlations to recursively simplify the optimization problem.

\subsection{\maxcut problem and its encoding}
A \maxcut problem instance is defined by a weighted, undirected graph $G = (V, E)$ with vertices $V = \{i\}$, edges $E = \{e\}$ and edge weights $w_{e}$.
We denote the number of vertices by $n = |V|$.
In $\text{\sc{MaxCut}}$, we are tasked with finding a subset of nodes that maximizes the weight of edges connecting the chosen node subset and its complement.
Formally, we want to find a node partition $W \subseteq V$ such that the edge set $\delta(W) \coloneqq \{ij \in E \mid i \in W, j \in V\setminus W\}$ maximizes its weight defined as $\sum_{e\in \delta (W)} w_e$.

Despite its simplicity, \maxcut is a popular example of a COP since any quadratic unconstrained binary optimization (QUBO) problem can be transformed into a \maxcut problem~\cite{ DESIMONE199071, Hammer1965}.
Furthermore, solving it is NP-hard~\cite{Karp1972}, and finding solutions for dense instances with hundreds of variables can already overstrain state-of-the-art algorithms~\cite{Charfreitag2022}.

The \maxcut problem can be formulated as maximizing an integer quadratic unconstrained cost function $C(\mathbf{x})$ in the form of
\begin{equation}
\label{eq: cost_function}
    C(\mathbf{x}) = \frac{1}{2} \sum_{ij \in E} w_{ij}(1-x_ix_j).
\end{equation}
Here, $\mathbf{x} \in \{-1, 1\}^n$, and $x_i$ indicates whether vertex $i$ is in the subset $W$ or not.
Furthermore, $w_{ij} \in \mathbb{R}$ represent the edge weights.

\subsection{Means of computing correlations}\label{subsec: means of computing correlations}
The core of our algorithm are correlations between decision variables of a \maxcut problem.
Each edge $ij\in E$ in the problem graph is assigned a correlation $b_{ij}\in [-1,1]$.
Ideally, we want $b_{ij}$ to represent the correlation between variables in high-quality solutions of the \maxcut problem.
In this case, a large negative (positive) correlation between two variables indicates that in high-quality solutions, these two variables mostly take opposite (equal) values.
In the language of $\text{\sc{MaxCut}}$, this translates to an edge predominantly being cut for negative correlations and not cut for positive correlations.

The shrinking algorithm works identically irrespective of the correlations that are used.
However, different means of computing correlations might be better suited for different instances.
Thus, in this paper, we compare three means of computing correlations: LP and SDP relaxations, and QAOA. 
All three run in polynomial time using classical or quantum resources.

\subsubsection{Linear programming (LP)}\label{sec: LP-correlations}
For a weighted, undirected graph $G = (V, E)$, we define an edge-incidence vector $\mathbf{y} \in \{0, 1\}^{|E|}$.
Here, a value of $y_e=1$ implies that edge $e$ is cut, while $y_e=0$ implies that edge $e$ is not cut.

With this notion, we model \maxcut as the integer linear program (cf.~\cite{barahona1988application, barahona1989experiments, junger2019odd, rehfeldt2023faster})
\begin{subequations}\label{eq: odd-cycle}
\begin{align}
    \begin{split}
    \label{eq: odd-cycle a}
    \max\quad &\sum_{e \in E} \omega_e y_e,
    \end{split} 
    \\[4pt]
    \begin{split}
    \label{eq: odd-cycle b}
    \text{s.t.:}\quad  &\sum_{e \in Q} y_e - \sum_{e \in C \backslash Q} y_e \leq |Q| - 1, \\[3pt]
      &\qquad \ \: |Q|\  \text{odd},\  \forall Q \subseteq C \ \text{cycle},
    \end{split}
    \\[4pt]
    \begin{split}
    \label{eq: odd-cycle c}
    &0 \leq y_e \leq 1,\ \forall e \in E, 
    \end{split}
    \\[4pt]
    \begin{split}
    \label{eq: odd-cycle d}
    &y_e \in \{0, 1\},\ \forall e \in E.
    \end{split}
\end{align}
\end{subequations}
Here, the \emph{odd cycle} inequalites~\eqref{eq: odd-cycle b} enforce the number of cut edges in any cycle to be even, which is sufficient for $y_e$ to define a cut.
While the \maxcut\ problem defined by Eqs.~\eqref{eq: odd-cycle a},~\eqref{eq: odd-cycle b},~\eqref{eq: odd-cycle d} cannot be solved efficiently in general, the linear relaxation defined by Eqs.~\eqref{eq: odd-cycle a}--\eqref{eq: odd-cycle c} can be solved in polynomial time and yields an upper bound on the maximum cut~\cite{Charfreitag2022}.
This also means that, if the relaxation returns an integer solution, the problem is solved to optimality.

Analogously to Ref.~\cite{wagner_enhancing_2023}, we compute correlations via the affine function
\begin{equation}\label{eq:lp_cor}
    b_e^\mathrm{LP} \coloneqq 1 - 2 \Tilde{y}_e \in [-1,1]
\end{equation}
where $\Tilde{\mathbf{y}} \in [0, 1]^{|E|}$ is an optimal solution to the linear program Eqs.~\eqref{eq: odd-cycle a}--\eqref{eq: odd-cycle c}.
Thus, we assign strong correlations to edges whose LP value is close to either $0$ or $1$.

\subsubsection{Semi-definite programming (SDP)}\label{sec: SDP correlations}
The calculation of semi-definite programming correlations is inspired by the seminal approximation algorithm introduced by Goemans and Williamson~\cite{goemans1995improved}.
The main idea underlying the Goemans-Williamson (GW) algorithm is that we replace the NP-hard integer quadratic problem formulation of \maxcut as introduced in Eq.~\eqref{eq: cost_function} by a relaxed version of the problem.
The relaxation admits a larger solution space that contains all possible solutions to the integer problem.
Thus, the optimal solution of the relaxed problem upper bounds the best solution of the integer model by design.
Concretely, we replace integer values with multi-dimensional vectors $\mathbf{v}_i \in \mathbb{R}^{n}$ with $\norm{\mathbf{v}_i}=1$, i.e., we expand the search space to the $(n-1)$-dimensional unit sphere $S_{n-1}$.
Using this, we can relax the maximization of Eq.~\eqref{eq: cost_function} to 
\begin{equation}\label{eq: GW Relaxed formulation}
    \begin{split}
        \max_{\{\mathbf{v}_1, \mathbf{v}_2, \dots\}} \quad &\frac{1}{2} \sum_{i < j} w_{ij} (1 - \mathbf{v}_i \cdot \mathbf{v}_j), \\
        \text{s.t.}\quad & \mathbf{v}_i \in S_{n-1} \quad \forall i \in V.
    \end{split}
\end{equation}
This model is equivalent to the semi-definite program
\begin{equation}
    \max_X \ \left\{\big\langle \frac{1}{4}L, X \big\rangle: \text{diag}(X) = \mathbf{{e}}, X  \succeq 0\right\},
\end{equation}
where $L \coloneqq \text{diag}(A \mathbf{{e}}) - A$ is the Laplacian of the problem graph, $A$ denotes the adjacency matrix, and $\mathbf{{e}}$ stands for the vector of all ones. 
For details see Refs.~\cite{goemans1995improved, anjos2011handbook}.
Here, $\langle a , b \rangle = \text{tr}(a b^\text{T})$ represents the Frobenius inner product of two matrices $a$, $b \in \mathbb{R}^{n \times m}$.
This generic semi-definite program can be solved in polynomial time by various solvers such as \texttt{cvxopt}, see, e.g.,~\cite{andersen2013cvxopt, yamashita2003implementation, anjos2011handbook}.

Furthermore, $X$ denotes the Gram matrix of $\{\mathbf{v}_i\}$.
To obtain the vectors $\{\mathbf{v}_i\}$, one needs to compute the Cholesky decomposition of the matrix $X$: $X = B^\text{T} B$, which has a computational complexity of $O(n^3)$.
In the resulting matrix $B$, the $i$th column corresponds to the vector $\mathbf{v}_i$.

We will consider two possibilities for calculating correlations based on SDP relaxations.
The first possibility, which we refer to as \emph{SDP correlations}, is more straightforward and can be directly computed from the vectors.
The second option -- \emph{GW correlations} -- requires additional evaluations of the $\text{\sc{MaxCut}}$ objective function.

\paragraph{SDP correlations}\label{parag: SDP correlations}
From  the vectors $\{\mathbf{v}_i\}$, we obtain the correlation of edge $ij \in E$ by evaluating the dot product
\begin{equation}\label{eq:sdp_cor}
    b_{ij}^\mathrm{SDP} = \mathbf{v}_i \cdot \mathbf{v}_j \in [-1,1],
\end{equation}
which measures the degree of (anti-)alignment of the vectors.
Thus, we assign a large absolute correlation if the vectors are approximately parallel or anti-parallel aligned.
The SDP correlations~\eqref{eq:sdp_cor} are motivated by the integer quadratic model~\eqref{eq: cost_function},
where an edge is cut if $x_i x_j = -1$, and not cut if $x_i x_j = +1$.
However, in contrast to the integer model~\eqref{eq: cost_function}, the SDP correlations can take continuous values between $-1$ and $+1$.
Furthermore, we can compute these correlations efficiently.

\paragraph{GW correlations}\label{parag: GW correlations}
Alternatively, we can obtain correlations by leveraging the hyperplane rounding procedure from the GW algorithm~\cite{goemans1995improved}.
The rounding procedure maps an optimal solution of~\eqref{eq: GW Relaxed formulation} to a \maxcut solution.
To this end, we first choose a random hyperplane, defined by a normal vector $\mathbf{r} \in S_{n-1}$.
Next, we divide the vertices $V$ into two partitions $S\subseteq V$ and $\overline{S} = V \, \backslash \, S$ using the criterion
\begin{equation}
    S = \{i \in V | \ \mathbf{v}_i \cdot \mathbf{r} \geq 0\}.
\end{equation}
Since we can efficiently evaluate the objective value of a given candidate cut,
we may try several (polynomially many) random hyperplanes and return the best solution found.
Ref.~\cite{goemans1995improved} shows that the cuts obtained this way are guaranteed to achieve an approximation ratio of at least $0.878$ on average.

For a given hyperplane that divides the vectors into two partitions, we define the \emph{GW correlations} as
\begin{equation}\label{eq:GW_cor}
    b_{ij}^\mathrm{GW} = 
    \begin{cases}
        \ \frac{1}{2} (\mathbf{v}_i \cdot \mathbf{v}_j + 1) & \text{if} \ i, \ j \ \text{are in the same partition,} \\[4pt]
        \ \frac{1}{2} (\mathbf{v}_i \cdot \mathbf{v}_j - 1) & \text{if} \ i, \ j \ \text{are in opposite partitions.}
    \end{cases}
\end{equation}
The GW correlations are designed to combine both the global information about the partition of the vertices
and the local information about the (anti-)parallel alignment of the vectors.

\begin{figure*}[]
  \centering
    \includegraphics[width=1\textwidth]{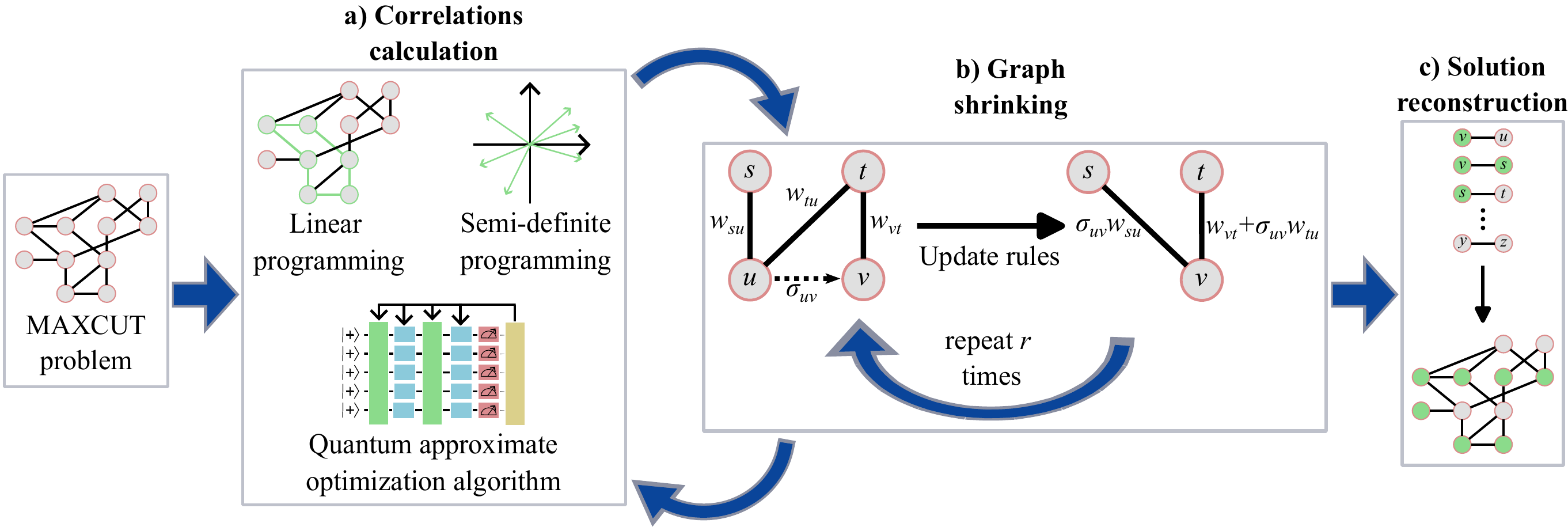}
    \caption{Depiction of the shrinking algorithm used in this work: First, the \maxcut problem instance is modeled according to the input requirement of the chosen routine for computing the correlations.
    In step a) we use the chosen routine to compute the correlations between variables.
    Then in b) the problem graph is shrunk to a smaller \maxcut problem by fixing a variable based on the calculated correlations and given update rules.
    The shrinking is repeated $r$ times before a new set of correlations is calculated for the simplified graph.
    The steps a) and b) are applied in an alternating manner until the problem is fully simplified.
    In the final step c) the solution to the original problem is reconstructed based on the fixed variables.}
    \label{Fig: Shrinking_algorithm_figure}
\end{figure*}

\subsubsection{QAOA}\label{subsec: QAOA introduction}
Finally, we compute correlations using QAOA. 
Here, we choose QAOA is a representative for various quantum routines.
Therefore, we remark that QAOA can readily be replaced by other quantum approaches, such as warm-start QAOA~\cite{egger_warm-starting_2021} or Quantum Annealing~\cite{mcgeoch2014adiabatic}.
We choose it because of its wide spread within research to achieve a proof of concept.
The goal of QAOA is to find an $\mathbf{x} \in \{-1,1\}^n$ that maximizes the cost function $C : \{-1, 1\}^n \rightarrow \mathbb{R}$ of an arbitrary integer quadratic unconstrained optimization problem.
Starting from the uniform superposition state $\ket{+}$ defined by
\begin{equation}
    \ket{+} = \frac{1}{\sqrt{2^n}} \sum_{\mathbf{x} \in \{-1, 1\}^n} \ket{\mathbf{x}},
\end{equation}
we apply a sequence of parametrized unitaries to prepare the state 
\begin{equation}
    \ket{\Psi (\boldsymbol{\beta},\boldsymbol{\gamma})} = e^{-i\beta_p H_M} e^{-i\gamma_p H_C} \cdots e^{-i\beta_1 H_M} e^{-i\gamma_1 H_C} \ket{+}.
\end{equation}
Here, $\boldsymbol{\beta} = (\beta_1, \dots, \beta_p)$, $\boldsymbol{\gamma} = (\gamma_1, \dots, \gamma_p)$ are real-valued parameters and $p\in \mathbb{N}$ is the depth.
$H_C$ denotes the cost Hamiltonian that encodes the optimization problem as
\begin{equation}
\label{eq: cost_hamiltonian_equation}
    H_C \ket{\mathbf{x}} = C(\mathbf{x}) \ket{\mathbf{x}} \quad \forall \mathbf{x} \in \{-1, 1\}^n,
\end{equation}
and $H_M$ is the so-called mixer Hamiltonian
\begin{equation}
    H_M = \sum_{i=1}^{n}X_i,
\end{equation}
where $X_i$ is the Pauli-X operator applied on qubit $i$.
We then use a classical subroutine that optimizes the parameters ($\boldsymbol{\beta}$, $\boldsymbol{\gamma}$) such that the expectation value
\begin{equation}
    F(\boldsymbol{\beta},\boldsymbol{\gamma}) = \bra{\Psi (\boldsymbol{\beta},\boldsymbol{\gamma})} H_C \ket{\Psi (\boldsymbol{\beta},\boldsymbol{\gamma})}
\end{equation}
is maximized.
Thus, we aim at preparing a superposition of high quality candidate solutions.

It follows from Eq.~\eqref{eq: cost_function} that, in the case of \maxcut, the cost Hamiltonian $H_C$ is given by
\begin{equation}\label{eq:HC}
    H_C = \frac{1}{2} \sum_{ij \in E} w_{ij}(I-Z_i Z_j),
\end{equation} 
where $I$ is the identity operator and $Z_i$ is the Pauli-Z operator acting on qubit $i$.
After maximizing the expectation value $F(\boldsymbol{\beta},\boldsymbol{\gamma})$,
we compute correlations $b_{ij}$ between nodes $i$ and $j$ via
\begin{equation}
    b_{ij}^\mathrm{QAOA} = \langle Z_iZ_j\rangle \equiv \bra{\Psi (\boldsymbol{\beta_{\text{opt}}},\boldsymbol{\gamma_{\text{opt}}})} Z_i Z_j \ket{\Psi (\boldsymbol{\beta_{\text{opt}}},\boldsymbol{\gamma_{\text{opt}}})},
\end{equation}
where $\boldsymbol{\beta_{\text{opt}}}$ and $\boldsymbol{\gamma_{\text{opt}}}$ are the optimized parameters.

In our experiments, we simulate QAOA with depths $p\in\{1, 2, 3\}$.
Ref.~\cite{ozaeta_expectation_2022} derives analytcial expressions for both the QAOA correlations $b_{ij}^\mathrm{QAOA}$ and the expectation value $F(\boldsymbol{\beta},\boldsymbol{\gamma})$ in the case $p=1$.
In the present work, we maximize the QAOA expectation value $F(\boldsymbol{\beta},\boldsymbol{\gamma})$ for $p=1$
by first applying a rough grid search over the parameter space of $\boldsymbol{\beta}, \boldsymbol{\gamma}$.
Then, we apply the gradient-based \texttt{BFGS} algorithm~\cite{wright2006numerical} with the best parameters from the grid-search as initial parameters.

For depths $p>1$, we use the \texttt{Qtensor} tensor network library~\cite{lykov2022tensor,lykov2021importance,lykov2022} to compute the expectation values and correlations.
The library maps the quantum circuits we want to evaluate to tensor networks.
This is done by interpreting a quantum state of $n$ qubits as a tensor from $(\mathbb{C}^2)^{\otimes n}$
and a quantum gate as a tensor with input and output indices for each qubit it acts on~\cite{boixo2018simulation}.
Here, an input index corresponds to the output index of the previous gate.
A contraction of the resulting tensor network leads to an exact simulation of the quantum circuit.
Because the required computational effort depends strongly on the order in which the indices are contracted, it is crucial to optimize the order of contraction~\cite{Markov_2008, Schutski2020}.
In this work, we use a method of finding a good contraction order based on a line graph representation of the tensor network as introduced in Refs.~\cite{lykov2021importance, lykov2022tensor}.

We use gradient ascent to find the optimal variational parameters $\boldsymbol{\beta_{\text{opt}}},\boldsymbol{\gamma_{\text{opt}}}$ that maximize
$F(\boldsymbol{\beta},\boldsymbol{\gamma})$.
To ensure the reliability of the optimization, it is essential to start gradient ascent from suitable initial parameters $\boldsymbol{\beta}_{\text{i}},\boldsymbol{\gamma}_{\text{i}}$~\cite{Lee_2021}.
In this work, we use the so-called \emph{fixed-angle conjectures} as initial parameters for the tensor network simulations guaranteeing a good performance~\cite{Wurtz2021}.
Although the conjectures are derived for $k$-regular graphs, we also use them as initial parameters for non-regular graphs with geometries similar to $k$-regular problems. 
These graphs arise from applying the shrinking steps (introduced in the next subsection) to $k$-regular graphs.
Our experiments show that QAOA, using parameters from the fixed-angle conjecture as initialization, produces output states with energies that are at least as good as those achieved with established parameter initializations such as using transition states~\cite{sack_recursive_2023} or interpolation methods~\cite{zhou_quantum_2020}. 

We would like to stress that even though QAOA can be run in polynomial time on a quantum computer (for fixed $p$),
it requires significant resources to simulate the algorithm using tensor networks. 
Moreover, the time required to simulate QAOA classically depends strongly on the \maxcut problem at hand. 
To keep computations tractable, we restrict the simulation to 3-regular graphs.

\subsection{Shrinking procedure}\label{subsec: shrinking procedure}

\maxcut is well-suited for shrinking algorithms since there is a natural way of reducing the problem such that the shrunk problem is still a valid \maxcut problem.
However, the shrinking procedure presented here can be extended to other problems as well~\cite{finzgar_quantum-informed_2024}.

The shrinking procedure used in this work is similar to RQAOA~\cite{bravyi_obstacles_2020, bravyi_hybrid_2022} and the algorithm introduced in Ref.~\cite{wagner_enhancing_2023}.
Fig.~\ref{Fig: Shrinking_algorithm_figure} presents a schematic of the proposed Algorithm~\ref{alg: shrinking algorithm}.

\begin{algorithm}
\caption{Shrinking algorithm}\label{alg: shrinking algorithm}
\hspace*{\algorithmicindent} \textbf{Input:} Graph $G = (V, E)$, \\
\hspace*{\algorithmicindent} \hspace*{0.975cm} Recalculation interval $r$, \\
\hspace*{\algorithmicindent} \hspace*{0.975cm} Steps $N$, \Comment{Number of shrinking steps} \\
\hspace*{\algorithmicindent} \textbf{Output:} Solution $S$. 
\begin{algorithmic}[1]

\State G' $\gets$ G
\State L $\gets$ $\{\}$ \Comment{Initialize an empty list for saving steps}
\State
\For{step in $\{0, ..., N-1\}$}
    \If{step mod r is 0} \Comment{Recalculate every r steps}
        \State Cor $\gets$ Calculate correlations for G'.\newline
        \hspace*{2.19cm}See section~\ref{subsec: means of computing correlations} and Fig.~\ref{Fig: Shrinking_algorithm_figure} (a).
        \State Cor $\gets$ Sort correlations by absolute value.
    \EndIf

    \State $(i, j)$, $\sigma_{ij}$ $\gets$ Pop first element of Cor.
    \If{i or j already collapsed onto other node}
        \State $(i, j)$, $\sigma_{ij} \gets$ Ensure consistency with G'.\newline
        \hspace*{3cm}Refer to text for further details.
    \EndIf
    \State G' $\gets$ shrink edge(G', (i,j), $\sigma_{ij}$) \Comment{See algorithm~\ref{alg: shrink edge} \\ \hspace*{5.96cm}and Fig.~\ref{Fig: Shrinking_algorithm_figure} (b)}
    \State Append shrinking step $\{(i, j), \sigma_{ij}\}$ to L.
\EndFor
\State
\State S' $\gets$ Solve \maxcut on shrunk graph G'.
\State S $\gets$ Reconstruct solution from solution $S'$ and history L. \hspace*{1.0cm} \Comment{See Fig~\ref{Fig: Shrinking_algorithm_figure} (c)}

\State \Return $S$ \Comment{Return Solution}
\end{algorithmic}
\end{algorithm}

It can be divided into three major steps.
We will give a brief overview before explaining the individual steps in detail.
For a given \maxcut problem, we first calculate correlations between the decision variables (step (a) in Fig.~\ref{Fig: Shrinking_algorithm_figure}).
Then, we use these correlations to successively reduce the number of nodes in the problem graph (step (b) in Fig.~\ref{Fig: Shrinking_algorithm_figure}).
To this end, we combine two nodes, generating a new \maxcut instance where the number of nodes is decreased by one.
The reduced problem is equivalent to the original problem with the additional constraint implied by the correlation.
Importantly, since the shrunk problem is again a \maxcut problem, we can calculate new correlations for the shrunk problem using the same method as for the original problem.
We re-calculate correlations after a given number $r$ of shrinking steps.
Finally, we recreate a solution to the original instance from a solution to the shrunk instance by undoing the shrinking (step (c) in Fig.~\ref{Fig: Shrinking_algorithm_figure}).
In the following, we describe the three steps in detail. 

\paragraph{Computing the correlations}\label{parag: computation of correlations}
The core of our algorithm is a method for computing correlations between decision variables of a given \maxcut instance.
We employ the methods presented in Section~\ref{subsec: means of computing correlations}.
However, the shrinking algorithm~\ref{alg: shrinking algorithm} can be used with any method for calculating correlations.
Importantly, its performance depends on the \emph{quality} of the employed correlations,
defined as the value of the best cut fulfilling these correlations.
In particular, the shrinking algorithm is guaranteed to return an optimal cut if the correlations are obtained from an optimal cut.

\paragraph{Shrinking}
Next, we present how correlations are used to guide the shrinking procedure.
A pseudocode representation of the following explanation can be found in algorithms~\ref{alg: shrinking algorithm} and~\ref{alg: shrink edge}.

\begin{algorithm}
\caption{Shrink edge}\label{alg: shrink edge}
\hspace*{\algorithmicindent} \textbf{Input:} Graph $G$, \\
\hspace*{\algorithmicindent} \hspace*{0.975cm} Correlation - edge $(i, j)$, sign $\sigma_{ij}$, \\
\hspace*{\algorithmicindent} \textbf{Output:} Graph G' with one less node than G.
\begin{algorithmic}[1]

\State E' $\gets$ get new edges from eq.~\eqref{eq: new nodes and edges}. 
\State Initialize empty weights $\omega'$.
\For{$(j, k)$ in E'} \Comment{Update edge weights}
    \If{(j, k) in E}
        \State $\omega'_{jk} \gets \omega_{jk} + \sigma_{ij} \omega_{ik}$
    \Else
        \State $\omega'_{jk} \gets \sigma_{ij} \omega_{ik}$
    \EndIf
\EndFor

\State G' $\gets$ Create graph from $\omega'$.
\State \Return G' \Comment{Return graph with one less node}
\end{algorithmic}
\end{algorithm}

After computing correlations, we choose the correlation with the largest absolute value, say $b_{ij}$, where ties are broken randomly.
The rationale is that we select edges first that have a strong tendency to be cut or not cut.
Then, we consider the sign $\sigma_{ij}$ of the correlation,
\begin{equation}
    \sigma_{ij} \coloneqq \text{sign}(b_{ij}).
\end{equation}
For positive (negative) $\sigma_{ij}$, the vertices $i$ and $j$
are fixed to lie in the same (opposite) partition, reducing the total number of vertices by one, see panel b) in Fig.~\ref{Fig: Shrinking_algorithm_figure}.
To ensure consistency, we update the weights of new graph $G'=(V', E')$ in the following way.
For a node $i \in V$, we define its neighborhood as $\mathcal{N}(i) \coloneqq \{ k \in V  \  |  \  ik \in E \}$.
For all nodes $k \in \mathcal{N}(i)$, we define new weights $\omega_{jk}'$ by
\begin{equation}
    w_{jk}' = 
    \begin{cases}
        \ \omega_{jk} + \sigma_{ij} \omega_{ik}, &\quad \text{if} \ jk \ \in E \\[4pt]
        \ \sigma_{ij} \omega_{ik}, &\quad \text{if} \ jk \ \notin E,
    \end{cases}
\end{equation}
while all other weights remain unchanged.
In comparison to the original graph $(G, E)$, the updated graph $(G', E')$ lacks node $i$ and some edges were merged,
\begin{equation}
    \label{eq: new nodes and edges}
    \begin{gathered}
        V' = V \, \backslash \, \{i\}  \\
        E' = \left(E \cup \{jk: \ k \in \mathcal{N}(i)\}\right) \, \backslash \, \left\{ik: \ k \in \mathcal{N}(i)\right\}.
    \end{gathered}
\end{equation}
An example of these update rules is given in Fig.~\ref{Fig: Shrinking_algorithm_figure} b) and the pseudocode for generating this updated graph can be found in Algorithm~\ref{alg: shrink edge}.
It is important to keep track of the imposed correlations in order to reconstruct the final solution at the end.

On a more technical note, it is possible that after several shrinking steps, a node $u$ belonging to the considered correlation $b_{su}$ does not exist anymore because it had already been combined with another node $v$.
This situation is shown in panel b) of Fig.~\ref{Fig: Shrinking_algorithm_figure}.
In this case, we simply interpret the correlation $b_{su}$ as the correlation $b_{sv}$.
Note however, that we must take care of the sign used in the previous step to ensure consistency.
To this end, we set $b_{sv} = \sigma_{uv} \cdot b_{su}$.
Furthermore, analogous rules apply if both nodes $i$ and $j$ have already been combined with other nodes.
For the special case that the two nodes $i$ and $j$ to be shrunk already belong to the same node $k$, the shrinking step is skipped, and the algorithm continues by using the next best correlation.
This skipping does not count as step in the procedure.

Since the shrinking step does not fix individual variables to be in a certain partition,
but rather fixes the relation between two variables, the shrunk problem remains a valid \maxcut problem.
This allows us to calculate correlations of the shrunk problem in the same manner as for the original graph.
Intuitively, correlations computed for the shrunk graph provide a better source of information for the next shrinking step than re-using correlations for the original graph.
Therefore, it makes sense to introduce recalculations of the correlations for the shrunk graph after a given interval of $r$ shrinking steps, as shown in Fig.~\ref{Fig: Shrinking_algorithm_figure} b) and Algorithm~\ref{alg: shrinking algorithm}.
With these newly obtained correlations, we shrink the reduced problem for another $r$ steps.
For correlations obtained by QAOA and a recalculation interval of $r=1$, the shrinking algorithm corresponds to the RQAOA algorithm from Refs.~\cite{bravyi_obstacles_2020, bravyi_hybrid_2022}.

In this work, we apply the shrinking procedure until the problem graph has only two nodes and is trivial to solve.
A potential extension of our algorithm performs the shrinking procedure until the problem graph has been reduced to a sufficiently small size such that an exact classical solver can find an optimal solution.

\paragraph{Reconstruction of the solution}
Finally, from a solution of the shrunk graph together with the history of all shrinking steps, we reconstruct a solution to the original graph, see Algorithm~\ref{alg: shrinking algorithm}.
Starting from a solution of the shrunk graph, we simply undo the performed shrinking steps.

\section{Results}
\label{sec: results}
Here, we present experimental results for the proposed shrinking algorithm using different methods for computing correlations, introduced in Sec.~\ref{subsec: means of computing correlations}.
We first introduce the considered \maxcut instances and performance metrics.
Then, we compare the performance of the shrinking algorithm with the different means of calculating correlations to the corresponding bare approximation algorithms.
Next, we investigate the influence of recalculating the correlations by studying the algorithm's performance at different recalculation intervals.
Finally, we employ tensor network methods to analyze the benefits of using deeper QAOA circuits to generate improved quantum correlations.

\subsection{Problem instances and performance metric}\label{subsec:problems}
We consider two families of graphs.
In most of our experiments, we solve \maxcut on Erd\H{o}s-R\'{e}nyi random graphs with one hundred nodes and different densities.
For a given number of nodes and density~$d$, an Erd\H{o}s-R\'{e}nyi graph is an undirected and unweighted graph where each
pair of nodes is connected by an edge with probability $d$.

Since simulating deep QAOA circuits requires computational resources that scale prohibitively with the graph density,
we use sparse graphs for the tensor network experiments.
Specifically, we use random 3-regular graphs with $50$ nodes.
This graph ensemble has been studied before in the context of QAOA for \maxcut\ problems~\cite{Wurtz2021, galda2021transferability, Zhou2020, Guerreschi_2019}
and has been used in hardware experiments because of the modest requirements on the qubit connectivity of quantum devices~\cite{Harrigan_2021, Lotshaw_2022}.

Throughout the results section we use the approximation ratio as the performance metric.
We define the approximation ratio as
\begin{equation}
\label{eq:approx_ratio}
    R_A \coloneqq \frac{S_A}{S_G}.
\end{equation}
Here, $S_A$ is the cut size obtained by the investigated algorithm and
$S_G$ is the cut size retrieved from Gurobi~\cite{gurobi}, a state-of-the-art solver.
For each instance, we run Gurobi on a single core (Dual AMD Rome 7742) and terminate the optimization when the objective value has not changed in an hour.
We emphasize that in this way, it is not guaranteed that Gurobi solves the problem instances to optimality.
However, we believe that the Gurobi benchmark is sufficiently strong for the purposes of this study.

\subsection{Comparison to the bare underlying algorithm}
\begin{figure}
    \centering
    \includegraphics[width=1\linewidth]{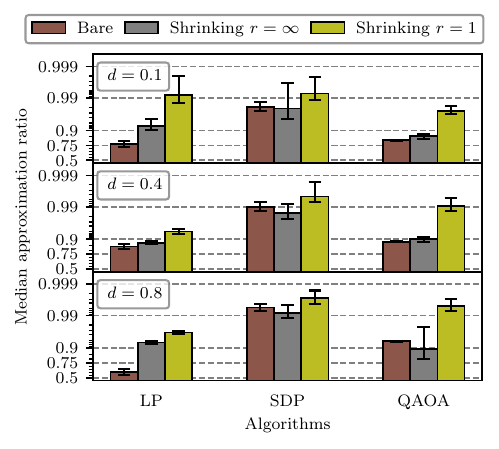}
    \caption{Median approximation ratio $R_A$ of the bare LP, GW and QAOA ($p=1$) algorithms as well as their shrinking counterparts with recalculation intervals $r=1$ and $r=\infty$.
    The algorithms are applied to 80 different randomly generated Erd\H{o}s-R\'{e}nyi graphs of size 100 for each of the densities $0.1$, $0.4$ and $0.8$.
    The lower and the upper error bars represent the first and third quartiles, respectively.}
    \label{fig: Comparison_algorithms}
\end{figure}

\begin{figure*}[]
  \centering
    \includegraphics[width=1.0\textwidth]{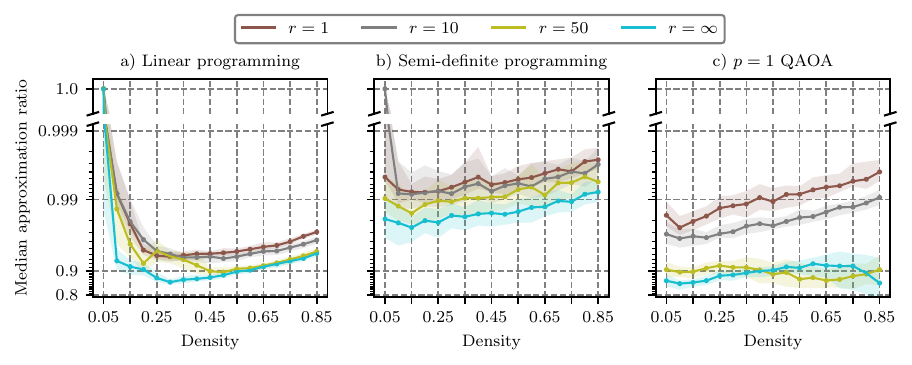}
    \caption{Median approximation ratio comparison of the shrinking algorithm with correlations from LP, SDP and QAOA ($p=1$) correlations.
    For each density between $0.05$ and $0.85$, eighty $100$-node Erd\H{o}s-R\'{e}nyi graphs were solved for different recalculation intervals $r = 1, \; 10, \; 50, \; \infty$.
    The shaded area represents the results between the first and third quartiles.}
    \label{Fig:Approximation_error_by_density}
\end{figure*}

We begin our analysis of the results by comparing the shrinking algorithm using various correlations to the solutions obtained by the bare underlying algorithms used to compute the correlations.
We first explain how the performance of the bare underlying algorithms is evaluated.

For the bare LP algorithm, we calculate the correlation for each edge in the manner described in Section~\ref{sec: LP-correlations}.
Then, we round these correlations to an integer solution by using a rounding heuristic based on a maximum-weight spanning-tree computation.
This is a well-established heuristic, previously used in Refs.~\cite{barahona1988application, liers2004computing, bonato2014lifting}.
For SDP, we use the standard Goemans-Williamson algorithm as the bare algorithm.
Here, we perform the hyperplane rounding (as described in Section~\ref{parag: GW correlations}) $15$ times and return the best solution.
For QAOA, the graphs are too large to perform a classical simulation of the quantum algorithm.
Efficient sampling from the output state of a QAOA $p=1$ circuit with one hundred qubits already exceeds the capabilities of classical hardware~\cite{FarhiHarrow}.
Therefore, we calculate the expectation value of the energy returned by $p=1$ QAOA using the analytical formulae from~\cite{ozaeta_expectation_2022}.

Fig.~\ref{fig: Comparison_algorithms} visualizes the results for comparing the shrinking algorithms to their bare counterparts.
Here, we generate $80$ random Erd\H{o}s-R\'{e}nyi graphs with one hundred nodes for each of the densities $d \in \{0.1, \; 0.4, \;0.8\}$.
The shrinking routine is applied with a recalculation interval of $r=1$ and $r=\infty$, i.e., without recalculations.

Starting with LP, we observe that the shrinking with no recalculations outperforms the spanning tree heuristic for all of the considered graph densities, even though the correlations are identical in both cases.
In addition, there is a clear improvement in performance when increasing the number of recalculations, that is, when decreasing $r$.
The same holds true for SDP and QAOA.
However, LP performs significantly better at lower densities than at high densities while SDP and QAOA show a roughly constant approximation ratio for all densities.

The bare GW algorithm and the SDP shrinking algorithm perform well across all densities.
The shrinking algorithm with recalculation interval $r=1$ performs best with median approximation ratios above $99\%$ for all densities.
However, the bare GW algorithm slightly edges out the shrinking algorithm with no recalculations.

Finally, for QAOA the shrinking algorithm with recalculation interval $r=1$ (equivalent to RQAOA) sees a significant improvement over the bare algorithm.
For all densities, the median approximation ratio increases roughly from around $90\%$ for the bare QAOA to approximately $99\%$ for the $r=1$ shrinking algorithm.
This behavior is in line with previous results from the literature~\cite{bravyi_obstacles_2020,bravyi_hybrid_2022}.
Interestingly, QAOA shows the largest relative improvement compared to the bare algorithm among the three different means of computing correlations.
The bare algorithm and the shrinking with no recalculations perform quite similarly.
However, the error bars indicate that the quality of the shrinking with no recalculations varies significantly more than the quality of the standard QAOA algorithm.

\subsection{Effect of the recalculation interval and problem density}\label{subsec:recalc and density}
In the previous section, we recalculated the correlations for the shrinking algorithm either after every shrinking step ($r = 1$) or not at all ($r = \infty$).
In this section, we vary the recalculation interval to investigate the influence on the performance.
Note that larger recalculation intervals reduce the computational cost.
To this end, we solve $80$ instances of Erd\H{o}s-R\'{e}nyi graphs for each density $d \in \{0.05,0.1,0.15,\dots,0.85\}$.
We use recalculation intervals $r \in \{1, \; 10, \; 50, \; \infty\}$.
The results for the shrinking algorithm using LP, SDP and $p=1$ QAOA correlations are shown in Fig.~\ref{Fig:Approximation_error_by_density}.

First, we focus on the effect of the recalculation interval before studying the impact of the problem density on the performance.
For all three means of computing correlations, we observe a clear trend that more recalculations improve the solutions.
However, in the case of LP correlations, for low densities, the $r=10$ shrinking algorithm slightly outperforms the $r=1$ variant.
While for LP and the SDP the performance differences between $r=10$ and $r=1$ are small, QAOA shows a significant increase from $r=10$ to $r=1$.
For SDP, LP and QAOA, the recalculation intervals $r=50$ and $r=\infty$ result in far worse approximation ratios than smaller recalculation intervals.
This highlights the importance of recalculating correlations.

Turning to the effect of problem density, the most striking dependence is observed for LP correlations.
There, the approximation ratio for low-density instances is significantly better than for any other type of correlation.
This is in agreement with the good performance of LP-based branch-and-bound algorithms for sparse \maxcut instances~\cite{Charfreitag2022,rehfeldt2023faster}.

A general feature in Figs.~\ref{Fig:Approximation_error_by_density} a) - c) is the systematic increase of the median approximation ratio for densities above $0.3$.
This increase is likely an artifact caused by a decreasing quality of the reference solutions obtained via Gurobi, rather than an improved performance of the shrinking algorithm.
While for sparse instances, Gurobi can certify that the returned solution is optimal, the same is not true for denser instances.
Thus, we believe that our choice of the benchmark metric causes the increasing trend in the approximation ratio observed across all correlation types.
However, our benchmark still allows us to compare between different correlation routines.

Thus, we observe that the performance of the LP shrinking algorithm decreases rapidly when the density increases from $0.05$.
In contrast, the SDP and QAOA correlations show to a more consistent performance of the shrinking algorithm across the entire range of densities.
Moreover, the SDP correlations clearly outperform $p=1$ QAOA for most densities, while the latter, in turn, outperforms LP correlations for densities above $0.2$.

\subsection{Performance of the GW correlations}\label{subsec: GW correlations performance}
After comparing SDP, LP and QAOA as means of computing correlations, we evaluate the GW correlations separately to ensure a fair comparison: 
While the previously analyzed strategies simply compute correlations using the problem graph, the algorithm for the GW correlations from paragraph~\ref{parag: GW correlations} evaluates the \maxcut objective function several times directly while searching for a good hyperplane.
Thus, it utilizes information not accessible to the other routines for computing correlations.
Nonetheless, these correlations can still be computed efficiently.

\begin{figure}[]
  \centering
    \includegraphics[width=\columnwidth]{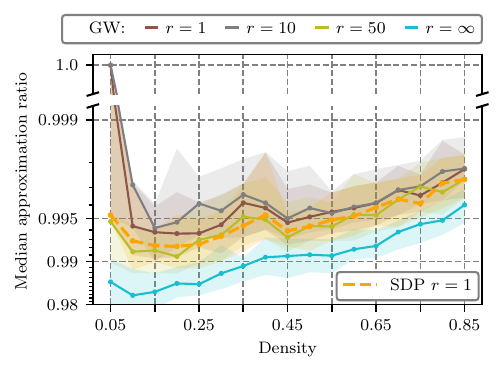}
    \caption{Median approximation ratio of the shrinking algorithm using the GW correlations for various recalculation intervals $r = 1, \; 10, \; 50, \; \infty$.
    For each density, $80$ $100$-node Erd\H{o}s-R\'{e}nyi graphs were solved.
    The shaded area represent the results between the first and third quartile.
    For comparison, the performance of the SDP correlations is also shown in this plot.}
    \label{Fig:Combi_results}
\end{figure}

The results for the shrinking algorithm using the GW correlations are shown in Fig.~\ref{Fig:Combi_results}.
During the rounding step, the algorithm chooses the best hyperplane out of $15$ tries.
In addition to the results for the shrinking procedure using GW correlations with recalculation intervals $r \in \{1, \; 10, \; 50, \; \infty\}$, we also plot the performance of the SDP correlations with a recalculation interval $r=1$.
This allows a direct comparison of shrinking via SDP correlations to the shrinking procedure using the GW correlations.
Importantly, because the signs of the GW correlations~\eqref{eq:GW_cor} coincide with the GW solution, the algorithm with $r=\infty$ returns the same solution as the GW rounding algorithm and thus inherits its performance guarantees.

Notably, the performance of the shrinking algorithm increases significantly by introducing recalculations and also clearly achieves better results than the Goemans-Williamson algorithm ($r = \infty$) for smaller recalculation intervals.
In general, the algorithm with GW correlations for the recalculation intervals $r \in \{1, \; 10\}$ also performs better than with the SDP correlations for $r=1$.
Thus, for densities above $0.1$, GW correlations yield the highest approximation ratio.
However, for densities below $0.1$, the LP correlations still perform best.

Furthermore, the GW correlations for $r=50$ have a similar performance as the SDP correlations for $r=1$.
From this, we conclude that the best approximation ratio for the SDP correlations can also be achieved by the GW correlations with less computational resources.
Finally, it is worth mentioning that GW correlations with a recalculation interval of $r=10$ yield a better median approximation ratio than with $r=1$ for all densities.
It seems that the additional information about a high-quality \maxcut solution, e.g. the well chosen hyperplane, is most beneficial when correlations are used for several shrinking steps without recalculations. Refer to section~\ref{sec: discussion} for a further discussion of this observation. 
This finding is particularly interesting since less computational resources yield a better result, in contrast to the analysis for the SDP correlations in the previous chapter.
There, the performance steadily increases with the number of recalculations.
As discussed more in detail in Section~\ref{sec: discussion}, this behavior could result from the additional information about the cut value to which only the GW correlations have access.

\subsection{Better quantum correlations lead to improved performance}
Next, we analyze the performance of the proposed shrinking algorithm informed by improved quantum correlations.
This is achieved by considering QAOA circuits at depths $p \in \{2, 3\}$.
The quality of QAOA is known to improve with increasing $p$~\cite{farhi2014quantum}.

We run the shrinking algorithm with QAOA depths $p \in \{1\, 2, 3\}$ and recalculation intervals $r \in \{1, \infty\}$.
For each pair of QAOA depth and recalculation interval, the algorithm is applied 5 times to each of the 25 problem instances on 3-regular graphs introduced in Sec.~\ref{subsec:problems}.
The simulations of QAOA are performed using tensor network methods as described in detail in Section~\ref{subsec: QAOA introduction}.

\begin{figure}[]
  \centering
    \includegraphics[width=\columnwidth]{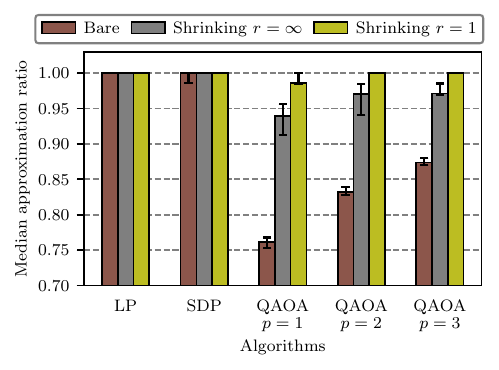}
    \caption{Median approximation ratio of the bare algorithms of LP, GW and QAOA (depths $p \in \{1, \;2, \;3\}$) as well as their shrinking counterparts with recalculation intervals $r=1$ and $r=\infty$ applied to 25 different random 3-regular \maxcut problem instances with 50 nodes.
    The bars represent the median value of applying the algorithms five times on each problem and the lower and upper error bars stand for the achieved first and third quartile of the individual approximation ratios, respectively.
    The approximation ratio is relative to optimum \maxcut solutions.}
    \label{Fig: Higher_depth_QAOA}
\end{figure}

Fig.~\ref{Fig: Higher_depth_QAOA} visualizes the results of our experiments, where the approximation ratio is computed with respect to the Gurobi solutions as defined in Eq.~\eqref{eq:approx_ratio}.
We note that in this experiment Gurobi was able to certify the optimality of the solutions.
For comparison, we also plot the results obtained via LP and SDP correlations.
We run the shrinking algorithm multiple times for each instance due to the random tie-breaking when choosing a correlation in a shrinking step.
This can lead to different final solutions.
This effect is particularly important for low depth QAOA, where the correlations depend only on the local neighborhood of a given edge.
For 3-regular graphs, there is a limited number of distinct local neighborhoods, which leads to many ties in the correlations~\cite{ozaeta_expectation_2022,Basso2022}.

Crucially, the results of Fig.~\ref{Fig: Higher_depth_QAOA} confirm our intuition that deeper QAOA circuits generate better correlations.
This intuition originates from the fact that the bare QAOA algorithm returns improved solutions as the depth $p$ increases.
Fig.~\ref{Fig: Higher_depth_QAOA} shows that the increased performance of bare QAOA for higher depths carries over to the shrinking algorithm, irrespective of the amount of recalculations performed.
Furthermore, we again observe that the performance of the shrinking algorithm improves significantly when recalculating the correlations after every shrinking step.
In fact, we are able to solve all of the considered graphs to optimality already at $p=2$ if the recalculation of correlations is performed after each shrinking step (yellow bars in Fig.~\ref{Fig: Higher_depth_QAOA}).
Nevertheless, we expect that for larger instances higher depths are required to obtain optimal solutions.
Finally, we note that the shrinking algorithm with no recalculations ($r=\infty$) on average outperforms the bare algorithm, even at higher depths.

The 3-regular graphs have a density of $d=0.06$. Thus, they correspond to the density regime where the LP based shrinking algorithm shows the best performance among the different algorithms (see Fig.~\ref{Fig:Approximation_error_by_density}).
Thus, all variants of LP return optimal solutions.
Furthermore, due to the smaller size of the problems the SDP algorithms likewise mostly return optimal solutions.
The SDP shrinking variant with $r=\infty$ solves more problem instances to optimality than the bare algorithm.
However, only the shrinking version with $r=1$ achieves optimal results for all \maxcut instances.

\section{Discussion}\label{sec: discussion}
In this work, we have extended and analyzed algorithms that solve a combinatorial optimization problem by recursively shrinking it.
We compared different methods for computing correlations, both quantum and classical, and thus provided classical benchmarks that quantum-informed shrinking algorithms will need to surpass in order to be useful in practice.
Importantly, our proposed shrinking procedure can be regarded as a standalone heuristic that improves the performance of the bare underlying algorithm.
We evaluated the algorithms through an extensive numerical analysis. 

% \paragraph{Extensions to existing algorithms} 
Interestingly, we observe that the shrinking algorithm not only increases the performance of QAOA, as already shown in~\cite{bravyi_obstacles_2020,bravyi_hybrid_2022}, but can also significantly improve upon the performance of classical algorithms like the Goemans-Williamson algorithm or the linear programming relaxation of $\text{\sc{MaxCut}}$.
Furthermore, the proposed shrinking procedure can serve as a heuristic for rounding LP or SDP relaxations to \maxcut solutions.
Our study indicates that the proposed algorithm delivers improved solutions compared to the well-known spanning-tree heuristic.

% \paragraph{Problem density} 
Applying the shrinking algorithm to Erd\H{o}s-R\'{e}nyi graphs of different densities shows that the LP correlations perform very well for low densities, with a rapid decrease in performance as the density increases.
This sensitivity of LP to the problem density can be explained by the relaxation model.
LP uses decision variables based on edges whereas the other relaxations use node-assignments as decision variables.
As a result, the number of variables for a constant number of nodes in the LP increases
proportionally to the density of the problem graph.
For the other relaxations, the number of variables is independent of the density.

The SDP and the QAOA correlations have a more constant performance where the approximation ratio slightly increases with increasing density.
As discussed in Section~\ref{sec: results}, this can be caused by both an increase in the quality of the shrinking algorithm solution or a decrease in the performance of the reference solution.

% \paragraph{Recalculations} 
Furthermore, we observed that more recalculations tend to yield better results.
However, this improved performance comes at the cost of an increase in the computational resources required to run the algorithm. Nevertheless, the overhead is at most linear even for recalculation interval $r=1$
Notably, the performance of the LP and SDP-informed shrinking algorithm only improves slightly when recalculating correlations in every step $(r=1)$ compared to every ten steps $(r=10)$.

% \paragraph{Trade-off for GW correlations} \label{parag: Trade-off for GW correlations}
The correlations from the GW-rounding perform best over the entire range of problem instances.
Interestingly, while there is a trend that the recalculations improve the performance, the best results with GW correlations are returned for a recalculation interval of $r = 10$.
In contrast, for the SDP correlations the performance increases monotonously with the number of recalculations, even though in both cases the correlations are based on the same algorithm.
Intuitively, the peak in performance at an intermediate value of $r=10$ can be understood by the competing effects of recalculations and GW rounding.
Our results indicate that increasing the number of recalculations improves the performance in general.
However, recalculating correlations also requires a new rounding hyperplane.
This new hyperplane might cause inconsistencies with previously performed shrinking steps.
We believe that this trade-off between the amount of recalculations and new hyperplanes is responsible for the observed best performance at $r=10$.

% \paragraph{Higher-order QAOA} 
Furthermore, we simulated higher-depth QAOA with up to three layers for three-regular graphs.
Our results confirmed that better quantum correlations improve the performance of the shrinking algorithm.
This indicates quantum hardware needs to improve before quantum algorithms can provide valuable approaches to combinatorial optimization.

\section{Outlook}\label{sec: outlook}
This study compares classical and quantum correlations for a recursive shrinking algorithm applied to \maxcut\ instances. 
Furthermore, we introduced a new GW-inspired approximation algorithm. 
To obtain quantum correlations, we used QAOA as a proof-of-concept quantum subroutine.
However, QAOA can be readily replaced by other quantum routines for computing variable correlations.
A further extension of our work would be to consider other optimization problems.
Moreover, there are several avenues towards increasing the algorithm's performance.

As discussed in~\cite{wagner_enhancing_2023}, one way to improve the performance of the shrinking algorithm is to avoid shrinking the problem until it becomes trivially to solve.
Instead, the algorithm shrinks the problem to a size where it can be solved to optimality by an exact solver.
This avoids mistakes during the last shrinking steps, leading to an improved solution.
Moreover, one could derive improved update rules or use backtracking techniques to rectify shrinking steps that caused a decrease in performance as proposed in~\cite{finzgar_quantum-informed_2024}.

Furthermore, an interesting property of the graph shrinking algorithm is that with each shrinking step both the graph connectivity and its edge weights change.
This is particularly interesting when considering persistency checks for \maxcut, which find provably optimal variable assignments~\cite{rehfeldt2023faster,glover2018logical}.
As the graph changes with each shrinking step, it is possible to conduct these checks at every shrinking step, even before calculating the correlations.
This can also only increase the performance of the algorithm if the checks are computed efficiently.

Another way of increasing the performance is updating the correlations if shrinking steps without recalculation are performed.
For example, if two edges are combined during a shrinking step, the new correlation assigned to this edge could be the mean of the correlations belonging to the original edges.

Furthermore, there are numerous other means of obtaining correlations beyond the ones discussed here.
For instance, much interest in quantum approaches to combinatorial optimization has been devoted to analog devices~\cite{kadowaki_quantum_1998, farhi_quantum_2001}, oftentimes paired with a classical optimizer to design the analog protocols~\cite{ebadi_quantum_2022, finzgar_designing_2024}.
Previously, shrinking algorithms using correlations from an analog device have been proposed in Ref.~\cite{finzgar_quantum-informed_2024}.
We believe that a systematic study of the performance of shrinking algorithms using quantum correlations from analog and gate-based quantum hardware compared to classical benchmarks could provide a useful stepping stone towards a better understanding of the utility of quantum computers for solving COPs.

\section*{Acknowledgments}

FW thanks Frauke Liers for fruitful discussions.
JRF acknowledges Libor Caha and Aron Kerschbaumer for insightful discussions.
VF and MP thank Johannes Frank for inspiring dialogues.
JRF and MP thank the BMW team for their support.
VF, FW, LP and CM are supported by the German Federal Ministry for Economic Affairs and Climate Action (BMWK), project QuaST.

\bibliographystyle{IEEEtran}
\bibliography{references.bib, more_references.bib}

\end{document}